\documentclass[cameraready]{Interspeech}

\usepackage{subcaption}
\usepackage{amsmath,bm}
\usepackage{multirow}
\usepackage{makecell}
\usepackage{caption}
\usepackage{subcaption}
\usepackage[table]{xcolor}
\newcommand{\tim}[1]{{\color{black}{#1}}}
\newcommand{\timtim}[1]{{\color{black}{#1}}}
\newcommand{\timtimtim}[1]{{\color{black}{#1}}}
\newcommand{\taozhong}[1]{{\color{black}{#1}}}

\title{Towards Personalized Federated Learning for Dysarthric Speech Recognition}

\author[affiliation={1}, , equalcontribution]{Tao}{Zhong}
\author[affiliation={2}, , equalcontribution, ]{Mengzhe}{Geng}
\author[affiliation={1}]{Jiajun}{Deng}
\author[affiliation={1}]{Shujie}{Hu}
\author[affiliation={1},,,correspondingauthor]{Xunying}{Liu}

\address{
    $^1$ The Chinese University of Hong Kong, China \\
    $^2$ National Research Council Canada, Canada
}

\email{tzhong@se.cuhk.edu.hk, Mengzhe.Geng@nrc-cnrc.gc.ca, 
jjdeng321@gmail.com,sjhu@se.cuhk.edu.hk, xyliu@se.cuhk.edu.hk}


\usepackage{comment}

\begin{document}

\maketitle

\begin{abstract}
Speech recognition is challenging for dysarthric speakers. While federated learning (FL)-based ASR can be an effective tool for protecting privacy, it suffers from heterogeneity issues caused by speaker variability. Forcing all speakers to share the same model components can be suboptimal under such heterogeneity, making personalization a promising direction; however, related research on dysarthric speech remains limited. To this end, this paper explores two aggregation strategies to achieve personalization, including the parameter-based averaging strategy and the embedding-based averaging strategy. Experiments on UASpeech and TORGO show that the proposed methods outperform the baseline regularized FedAvg by statistically significant WER reductions of up to 0.99\% absolute (3.15\% relative) on UASpeech and 0.56\% absolute (4.73\% relative) on TORGO, respectively.
\end{abstract}

\section{Introduction}
\label{sec:intro} 
\tim{Despite decades of progress in automatic speech recognition (ASR) systems for typical speech~\cite{dong2018speech,gulati20_interspeech}, accurate recognition of dysarthric speech remains challenging to date~\cite{geng2020investigation,geng2022speaker,yue2022acoustic,baskar2022speaker,10584335,hsu2024cluster,wang2025phone,jin2023personalized,wang2024enhancing}.}  
\tim{Dysarthric speech presents multi-fold challenges for current deep learning-based ASR systems primarily targeting healthy users, including}:
\textbf{1) substantial mismatch} between this data and normal voices; \textbf{2) data scarcity} resulting from the difficulties associated with data collection, frequently constrained by speakers' mobility limitations~\cite{10584335}; 
and \textbf{3) large speaker heterogeneity}
\tim{, where variations in accent and gender are further exacerbated by speech impairments}. Given the high priority of \tim{privacy concerns in healthcare among dysarthric speakers}, there has been a growing shift towards decentralized
training over centralized approaches~\cite{williamson2024balancing}.

\tim{Federated learning (FL)\cite{pmlr-v54-mcmahan17a} offers a powerful solution to data privacy concerns by enabling collaborative model training across decentralized datasets.} 
\tim{In recent years, FL has been widely adopted for speech-related tasks targeting the healthy population, such as speaker verification~\cite{leroy2019federated,hard2020training}, speech emotion identification~\cite{Granqvist2020,zhang2024stealthy}, keyword spotting~\cite{latif2020federated,feng2022semi}, and speech recognition~\cite{dimitriadis2020federated,guliani2021training,yu2021federated,cui2021federated,nandury2021cross,gao2022end,azam2023importance,azam2023federated,du2024communication}.}
\tim{Despite these advancements, the application of FL to speech-related tasks in healthcare remains limited. Most existing studies focus on the detection of Alzheimer's disease~\cite{meerza2022fair} or Parkinson's disease~\cite{arasteh2023federated}, with far fewer addressing speech recognition~\cite{hsu2024cluster,zhong25_interspeech}.}

\tim{Model aggregation lies at the core of federated learning, where local models are integrated into a global model. Widely adopted methods like FedAvg~\cite{pmlr-v54-mcmahan17a} utilize a quantity-based averaging strategy, weighting client model updates by the proportion of data samples contributed by each client. For speech-related tasks, integration of task-specific knowledge has been explored in~\cite{gao2022end}, where word error rate (WER) based aggregation is proposed.} \tim{However, the \textbf{substantial heterogeneity among dysarthric speakers} suggests that a single global model may not effectively model all speakers}, and adding personalization~\cite {du2024communication, farahani2023toward,hsu2024cluster,10389738} is a direction to tackle the above issues and the related research on dysarthric speech is limited and needs exploration.

To this end, this paper proposes two \tim{novel} model aggregation strategies \tim{targeting personalized federated learning for dysarthric speech recognition.} \tim{Specifically, the model is divided into a speaker-independent (SI) component and a speaker-dependent (SD) component.} \tim{For the \textbf{SI part}, standard FedAvg is employed for averaging. For the \textbf{SD part}, two similarity-based averaging strategies are introduced: \textbf{(1) parameter-based averaging}, which computes speaker similarities based on model parameters; and \textbf{(2) embedding-based averaging}, which calculates speaker similarities using output embeddings from the SI component. In order to improve privacy protection, we calculate client embeddings using a random subset of private data in each communication. These speaker similarities are then used to guide the averaging process of the SD model component, enabling personalized models tailored to individual dysarthric speakers.} Performance evaluation is conducted on two benchmark healthcare datasets: 1) UASpeech~\cite{kim2008dysarthric} dysarthric speech corpus; and 2) TORGO~\cite{rudzicz2012torgo} dysarthric speech corpus.

The main contributions of this paper are as follows:


\timtim{\noindent \textbf{1)} To the best of our knowledge, this is the first work on personalized federated learning for dysarthric speech recognition. We propose two similarity-aware aggregation strategies that personalize FedAvg using inter-speaker similarity. Prior personalized FL studies focus on normal speech~\cite{du2024communication,10389738,NEURIPS2020_24389bfe}. Compared with~\cite{du2024communication}, which personalizes via k-nearest neighbors, we achieve personalization through similarity-weighted aggregation. Moreover, we build on the large pretrained HuBERT model rather than a Conformer backbone as in~\cite{10389738}.} \taozhong{Unlike \cite{NEURIPS2020_24389bfe}, which risks poor initialization due to the averaging of conflicting gradients from speaker heterogeneity and relies on isolated client fine-tuning, our approach actively pulls updates toward clinically relevant speaker neighbors to minimize the negative interference of unrelated dysarthric speech patterns.}


\timtim{\noindent \textbf{2)} Experiments on the benchmark UASpeech and TORGO datasets show statistically significant WER reductions of up to 0.99\% absolute (3.15\% relative) and 0.56\% absolute (4.73\% relative), compared with a regularized FedAvg baseline.}

\tim{The rest of the paper is organized as follows. Section~\ref{sec:fedasr} describes the \timtimtim{FL-based} ASR systems. The two aggregation-based personalized FL techniques, i.e., parameter-based averaging and embedding-based averaging, are detailed in Section~\ref{sec:person}. Section~\ref{sec:exp} presents experiments on UASpeech and TORGO. Section~\ref{sec:conclusion} concludes the paper and discusses future work.}

\section{Federated Learning Based ASR}\label{sec:fedasr}

\begin{figure}[htbp]
   \centering
\includegraphics[width=0.43\textwidth]{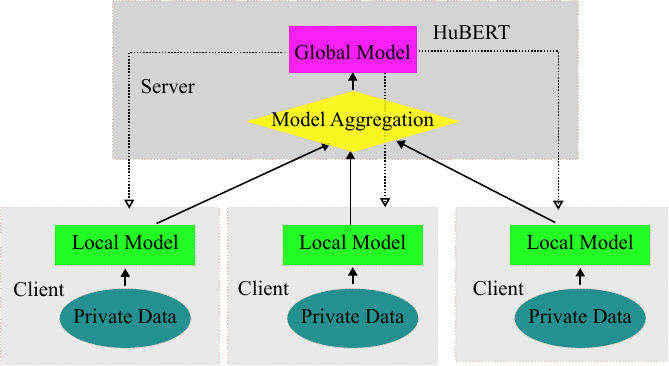}
   \vspace{-0.2cm}
    \caption{Illustration of the federated learning based HuBERT ASR system. Each communication round includes: \textbf{(a)} aggregating the parameters of locally trained client-side models into a global model (solid line) and \textbf{(b)} redistributing the global model to clients (dashed line).}
\label{fig:fed_learning}
\end{figure}

In contrast to standard distributed training~\cite{verbraeken2020survey}, federated learning (FL) \tim{adheres to privacy requirements by keeping data on local devices} (shown in Fig.~\ref{fig:fed_learning}). During each communication round, local updates are \tim{transmitted} to the central server and aggregated into a global model\tim{, which is then distributed back to each client.} Using local model updates instead of sharing raw gradients significantly enhances privacy in federated learning.

\begin{figure}[htbp]
   \centering
   \begin{subfigure}[ht]{0.45\textwidth}\includegraphics[width=\textwidth]{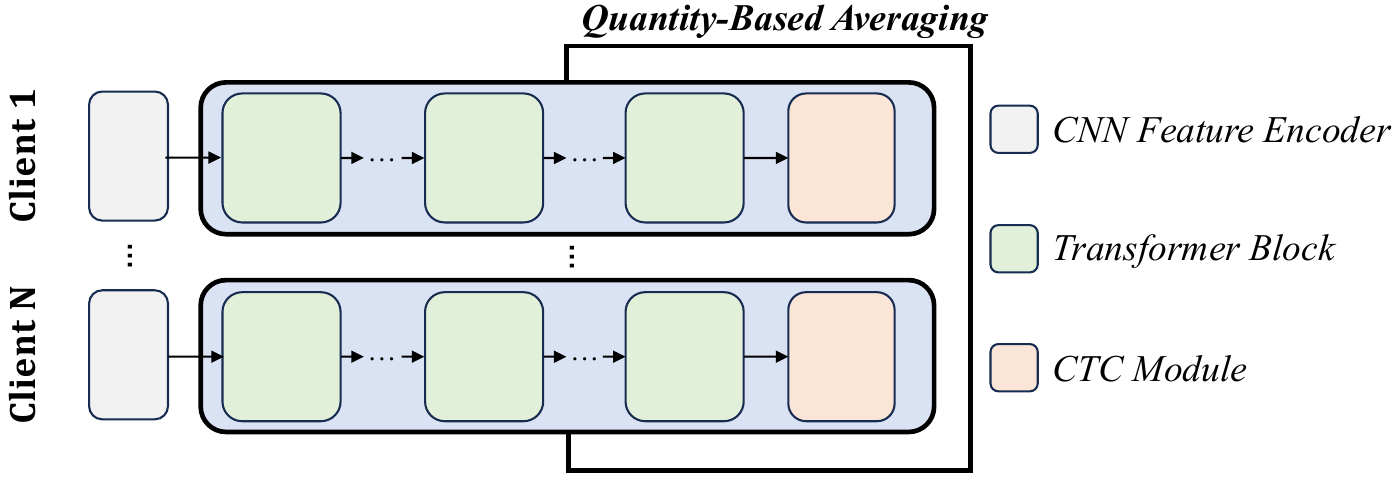}
   \caption{\textbf{FedAvg}, which averages all parts of client models into one global model.}
   \end{subfigure}
   \begin{subfigure}[ht]{0.45\textwidth}
    \includegraphics[width=\textwidth]{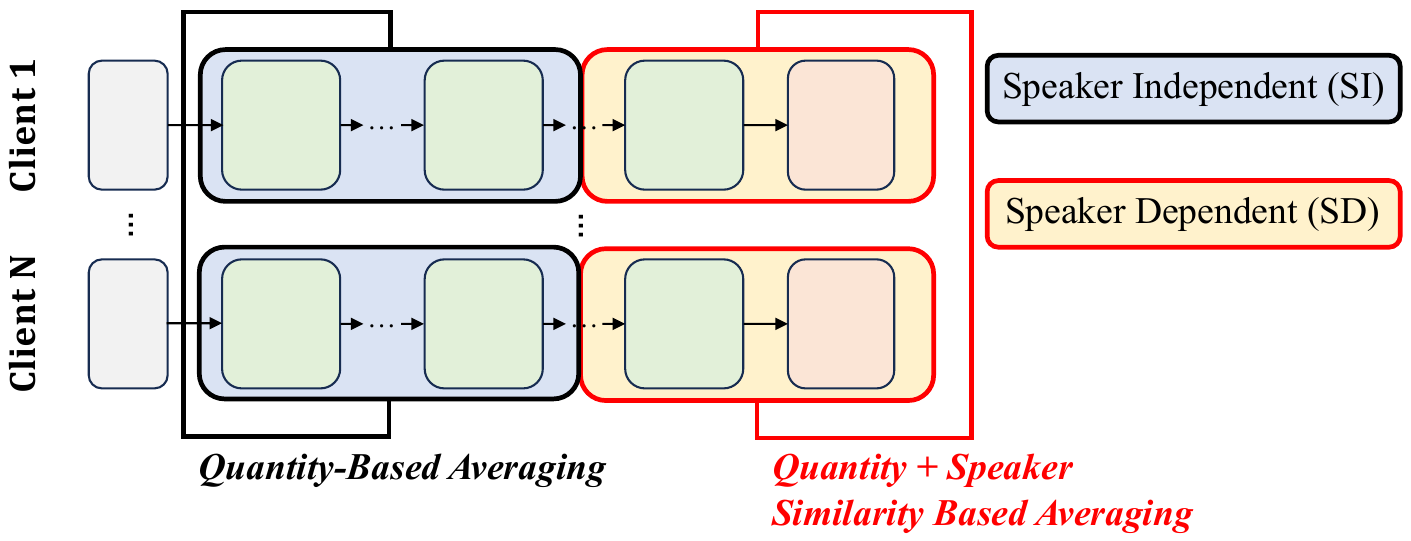}
    \caption{\textbf{Our proposed parameter-based averaging and embedding-based averaging}, where the SI and SD parts of client models are averaged separately.}
   \end{subfigure}
   \vspace{-0.3cm}
    \caption{Illustration of the standard FedAvg and our proposed approaches. FedAvg uses an SI quantity-based averaging. In contrast, our approaches combine SI quantity-based averaging with SD quantity + speaker similarity based averaging, where the speaker similarities are computed from model parameters (Sec.~\ref{ssec:para}) or embeddings(Sec.~\ref{ssec:emb})\protect\footnotemark.}
\label{fig:fedagg}
\end{figure}

As shown in Fig.~\ref{fig:fedagg}(a), FedAvg~\cite{pmlr-v54-mcmahan17a} is \tim{a foundational} \tim{FL} algorithm \tim{using} \tim{a} quantity-based \tim{averaging} strategy, where the weight is proportional to the number of data samples in each client: 

\vspace{-0.4cm}
\begin{equation}
\overline{\bm{w}}^{t} \leftarrow \sum_j^N \frac{|D_j|}{\sum_m^N {|D_m|}} \bm{w}_j^{t} 
\end{equation}\label{eqn:fedavg}
\vspace{-0.1cm}

\noindent \tim{where $\overline{\bm{w}}^{t}$ represents the averaged global model parameter in the $t^{th}$ communication round, $\bm{w}_j^{t}$ denotes the local model parameter from the $j^{th}$ client, $|D_j|$ is the number of data samples on the $j^{th}$ client, and $N$ is the total number of clients.}

\tim{To further tackle data heterogeneity across clients, multiple regularization techniques are explored in~\cite{zhong25_interspeech}, including parameter-based, embedding-based, and loss-based regularizations. The parameter-based approach seeks to align the local model's parameters with the global model's parameters. Similarly, embedding-based regularization focuses on aligning intermediate embeddings between the two models. Finally, loss-based regularization employs the KL divergence to align the output distributions of the local and global models.}

\vspace{-0.2cm}

\footnotetext{The SI components are positioned below the SD components to ensure that embedding similarity is evaluated within a unified vector encoding space defined by the SI components shared across all speakers (devices).}

\vspace{-0.1cm}
\section{Personalized federated learning based ASR}
\label{sec:person}

\subsection{\tim{Parameter-Based Averaging}}
\label{ssec:para}

\vspace{-0.2cm}
\tim{Fig.~\ref{fig:fedagg}(b) illustrates the aggregation process for the SD part in federated learning using parameter-based \tim{averaging}. The trainable components, excluding the fixed CNN feature encoder, are divided into two parts: the SI part, which is shared across all speakers, and the SD part, which is specific to each speaker. The method comprises two main steps:}

\noindent\tim{\textbf{Step 1:} The SD part is fixed while each client trains the SI part. The trained SI parts are then sent to the server, where they are aggregated according to Equation~(1).} \\
\noindent\tim{\textbf{Step 2:} The SI part is fixed while each client trains the SD part. The trained SD parts are subsequently sent to the server to calculate the speaker similarities, given as:}
\vspace{-0.2cm}
\begin{equation}
S_{i,j,l}^{t}\!\leftarrow\frac{\exp(cos\_sim((\bm{w}_{i,l}^t\!-\bm{w}_{i,l}^0), (\bm{w}_{j,l}^t\!-\bm{w}_{j,l}^0))}{\sum_k \exp(cos\_sim((\bm{w}_{i,l}^t\!-\bm{w}_{i,l}^0),(\bm{w}^t_{k,l}\!-\bm{w}^0_{k,l}))}
\end{equation}\label{eqn:para_sim}
\vspace{-0.2cm}

\noindent where $S_{i,j,l}^{t}$ \tim{represents} the similarity between the $i^{th}$ client and the $j^{th}$ client \tim{for} the $l^{th}$ layer \tim{during} the $t^{th}$ communication round. $\bm{w}^0_{i,l}$ is the initial weight of the $l^{th}$ layer in the $i^{th}$ client. $\bm{w}^t_{i,l}$ is the \tim{updated} weight in the $t^{th}$ communication round\footnote{\tim{The SD superscripts  are omitted in Equation~(2) for simplicity.}}. 
\tim{$cos\_sim(\cdot)$ denotes the cosine similarity.} \tim{The averaging of the SD parts is then performed as:}

\vspace{-0.6cm}
\begin{equation}
\overline{\bm{w}}_{i,l}^{t, SD}\! \leftarrow\! \sum_j^N \left((1\!-\!\beta) \cdot \frac{|D_j|}{\sum_m^N {|D_m|}}\! +\! \beta \cdot S_{i,j,l}^{t}\right) \bm{w}_{j,l}^{t, SD}\label{eqn:para_agg}
\end{equation}
\vspace{-0.2cm}

\noindent \tim{where }$\overline{\bm{w}}_{i,l}^{t, SD}$ is the aggregated weight of the $l^{th}$ layer in the SD part \tim{for} the $i^{th}$ client \tim{during} the $t^{th}$ communication round. $\beta$ is the trade-off weight to balance the quantity-based and similarity-based \tim{averaging}. $N$ is the number of clients.

\timtimtim{The combination of steps 1 and 2 is treated as a single communication round, since the communication cost is comparable to that of a single round in standard FedAvg (the fixed parts are not sent to the server).} Besides, the similarity computation is performed after local training, and the computational cost is small relative to the local training.

\subsection{\tim{Embedding-Based \tim{Averaging}}}
\label{ssec:emb}



The aggregation of the SD part with embedding-based averaging is shown in Fig.~\ref{fig:fedagg}(b). The method follows three steps: 

\noindent \textbf{Step 1:} The SI part is aggregated according to Equation~(1).

\noindent \textbf{Step 2:} In the $i^{th}$ client, we obtain the output from the SI part for each data sample. \timtimtim{Then, mean pooling is applied to the output along the sequence dimension, and the resulting pooled representation is averaged over the data to obtain the client embedding $\bm{e}_i^t$ in the $t^{th}$ communication round.}

\noindent \textbf{Step 3:} The SI part is fixed, and each client trains the SD part before sending the SD part to the server to calculate the speaker similarities and conduct the aggregation:
\vspace{-0.2cm}
\begin{equation}
S_{i,j}^{t}\leftarrow\frac{\exp(cos\_sim(\bm{e}_i^t, \bm{e}_j^t))}{\sum_k \exp(cos\_sim(\bm{e}_i^t,\bm{e}_k^t))}
\end{equation}\label{eqn:emb_sim}
\vspace{-0.3cm}

\noindent where $S_{i,j}^{t}$ \tim{represents} the similarity between the $i^{th}$ client and the $j^{th}$ client \tim{during} the $t^{th}$ communication round. $\bm{e}^t_{i}$ is the client embedding in the $i^{th}$ client \tim{during} the $t^{th}$ communication round. 
\tim{$cos\_sim(\cdot)$ denotes the cosine similarity.} \tim{The averaging of the SD parts is then performed as:}
\vspace{-0.2cm}
\begin{equation}
\overline{\bm{w}}_i^{t, SD}\! \leftarrow \!\sum_j^N \left((1-\beta) \cdot \frac{|D_j|}{\sum_m^N {|D_m|}}\! + \!\beta \cdot S_{i,j}^{t}\right) \bm{w}_j^{t, SD}\label{eqn:emb_agg}
\vspace{-0.2cm}
\end{equation}
\noindent \tim{where }$\overline{\bm{w}}_{i}^{t, SD}$ is the aggregated weight in the SD part for the $i^{th}$ client (or speaker) during the $t^{th}$ communication round.

Motivated by~\cite{NEURIPS2018_3b5020bb}, in order to enhance privacy, we sample 20\% of the data within each private client to calculate client embeddings in each communication round.

We treat the combination of step 1, step 2, and step 3 as one communication round since the communication cost is comparable with that of one communication round in FedAvg described in Section~\ref{sec:fedasr} (the fixed parts will not be sent to the server, and the size of the embeddings is very small). Besides, the computation of the similarity is conducted after local training, and the computation cost is small (the size of the embeddings is small) compared with the local training. 

\vspace{-2mm}
\section{Experiments and Results}
\label{sec:exp}
\subsection{Task Description}
\textbf{The English UASpeech} corpus~\cite{kim2008dysarthric} is the largest publicly available and widely used dataset for dysarthric speech recognition. It comprises an isolated word recognition task with approximately 103 hours of speech data from 29 speakers, among whom 16 are dysarthric speakers and 13 are healthy control speakers. The dataset includes 155 common words and 300 uncommon words and is further divided into three blocks B1, B2 and B3. The same 155 common words are used across all blocks, while the 300 uncommon words differ between blocks. In our experiments, we focus \textbf{on the 16 dysarthric speakers} and exclude the healthy control speakers. B1 and B3 are used for training, while B2 is used for evaluation. After removing silence, the training set and the test set contain 17.8 hours (52785 utterances) and 9 hours of audio (26520 utterances) in total, respectively. Speech intelligibility assessment is available for the 16 dysarthric speakers, divided into four groups: ``very low'' (VL), ``low'' (L), ``mid'' (M) and ``high'' (H).

\noindent \textbf{The English TORGO} corpus~\cite{rudzicz2012torgo} is a dysarthric speech dataset containing 8 dysarthric and 7 control healthy speakers with 13.5 hours of speech (16433 utterances). The entire TORGO dataset contains 1573 distinct words. Similar to the setting in UASpeech, in our experiments, we focus \textbf{on the 8 dysarthric speakers} and exclude the healthy control speakers, and we put two-thirds of the dysarthric speakers’ data into the training set. The remaining one-third of the dysarthric speech is used for performance evaluation\taozhong{~\cite{10584335}}. After the removal of excessive silence and data augmentation, the augmented training and test sets respectively contain 15 hours (25580 utterances) and 1 hour (1892 utterances) of speech. These 8 dysarthric speakers can be divided into 3 subgroups, namely, “severe”, “moderate”, and “mild” intelligibility subgroups. 

\vspace{-2mm}
\subsection{Experiment Setup}
\textbf{Model configuration:} HuBERT~\cite{hsu2021hubert} model\footnote{https://huggingface.co/facebook/hubert-large-ls960-ft} fine-tuned on 960 hours of Librispeech~\cite{panayotov2015librispeech} is used in our experiment. The loss is the Connectionist Temporal Classification (CTC)~\cite{graves2006connectionist}. The model consists of the CNN feature encoder (fixed), a stack of 24 Transformer blocks, and the final fully connected layer for CTC training.  Each client performs local training for one epoch before sending the local model to the server. The number of communication rounds is set to 100. Experiments are conducted using two Nvidia A40 GPUs. A matched pairs sentence-segment word error (MAPSSWE)~\cite{bisani2004bootstrap} based statistical significance test is performed with a significance level of $\alpha = 0.05$.

\noindent \textbf{Client partitioning:} For the UASpeech dysarthric speech task, the training data of each of the 16 dysarthric speakers is assigned to a separate client, resulting in a total of 16 clients, each containing 0.71 to 1.54 hours of audio (1785 to 3570 utterances). For the TORGO dysarthric speech task, the training data of each of the 8 dysarthric speakers is assigned to a separate client, resulting in a total of 8 clients, each containing 0.54 to 2.41 hours of audio (1064 to 4935 utterances).

\noindent \textbf{Baselines:}
\begin{itemize} 
\item \textbf{Centralized learning}: In stage a, we treat the whole model as the SI part, and we train the model to converge with all the data in the server. In stage b, we add adaptors (each adaptor is composed of two fully connected layers with ReLU between them, and it is placed after the fully connected layer of each transformer block)  as the SD part and the model obtained in stage a as the SI part and fine-tune the model~\footnote{Treating  $7^{th}$-$24^{th}$ transformer layers and the last fully connected layer (CTC module) as the SD part will result in the GPU OOM.} (Sys. 0a and 0b in Table~\ref{tab:ua_aggregation} and~\ref{tab:torgo_aggregation}).

\item \textbf{Regularized FedAvg}~\cite{zhong25_interspeech}: FedAvg algorithm (quantity-based averaging) regularized by parameter, embedding, and loss-based 
regularization\footnote{The penalty weight is set as 0.001, 0.001, 0.01 for parameter, embedding, and loss-based regularization.} (introduced in Section~\ref{sec:fedasr}) (Sys. 1 in Table~\ref{tab:ua_aggregation} and~\ref{tab:torgo_aggregation}). \taozhong{Regularized FedAvg incorporates the mechanics of FedProx~\cite{MLSYS2020_1f5fe839} (parameter-based) while adding additional regularization approaches to achieve superior results.}
\end{itemize}

\noindent \textbf{Proposed methods:}
\begin{itemize}
\item \textbf{Federated learning with parameter-based \tim{averaging}}: As shown in \timtimtim{Section~\ref {ssec:para}}, the SI part is aggregated in step 1\footnote{\label{note1}We apply parameter and embedding-based regularization to the SI part in step 1. We ignore the loss-based regularization since it needs the output logits for regularization, and the model has several outputs.  The penalty weight is set as 0.001, 0.001 for \timtimtim{the} parameter, embedding-based regularization.}, and the SD part is aggregated in step 2\footnote{\label{note2}The trade-off $\beta$ is set as 0.8 for UASpeech and 0.6 for TORGO.} (Sys. 2-5 in Table~\ref{tab:ua_aggregation}, \ref{tab:torgo_aggregation}). 

\item \textbf{Federated learning with embedding-based \tim{averaging}}: As shown in Section~\ref{ssec:emb}, the SI part is aggregated in step 1\footref{note1}, and the SD part is aggregated in step 3\footref{note2}(Sys. 6-9 in Table~\ref{tab:ua_aggregation}, \ref{tab:torgo_aggregation}). 
\end{itemize}

\subsection{Results Analysis}
\vspace{-2mm}
\noindent \textbf{Experiments on the UASpeech corpus}: Table~\ref{tab:ua_aggregation} compares
parameter-based, embedding-based \tim{averaging} for federated learning. Several trends can be
observed: \textbf{1)} Both these aggregations lead to 
performance improvements over the baseline regularized FedAvg system (Sys.2-9 vs. Sys.1).  \textbf{2)} By applying parameter-based (Sys.2-5) or embedding-based \tim{averaging} (Sys.6-9) with different SD parts, with
statistically significant overall WER reductions of up to 0.94\% abs. (2.99\% rel., Sys.3 vs. Sys.1) and 0.99\% abs. (3.15\% rel., Sys.7 vs. Sys.1) over the baseline regularized FedAvg system,
respectively. \textbf{3)} Combining both leads to
further performance improvement\footnote{trade-off weight is set as 0.2, 0.3,  0.5 for quantity-based, parameter-based and embedding-based \tim{averaging} strategy.}(Sys.10). \textbf{4)} The improvement brought by the personalization in FL training (1.02\% abs., Sys.10 vs Sys.1) is comparable to that obtained in centralized training settings (1.20\% abs., Sys.0b vs Sys.0a). \taozhong{\textbf{5)} Very Low (VL) users face the greatest barriers in ASR. Our method yields a 2.47\% abs. WER reduction for this group. This specifically addresses severe acoustic deviations where the standard global model performs poorly.}

\vspace{-0.3cm}
\begin{table}[!htbp]
    \centering
    \caption{Performance of parameter (``para.''), embedding (``embed.'')-based \tim{averaging}  for federated learning on \textbf{UASpeech}. ``$1$:$3$'' represents that $1^{st}$-$3^{rd}$ transformer layers are chosen as the SI part, and the rest as the SD part ($4^{th}$-$24^{th}$ transformer layers and the last fully connected layer). ``$1$:$6$'', ``$1$:$12$'', ``$1$:$18$'' have similar meanings.``VL'', ``L'', ``M'' and ``H'' are ``very low'', ``low'', ``mid'' and ``high'' speech intelligibility. $^{\dag}$ denotes a statistically significant improvement $(\alpha = 0.05)$ obtained over the regularized FedAvg system (Sys.1).}
    \label{tab:ua_aggregation}
    \vspace{-0.3cm}
    \setlength{\tabcolsep}{1pt} 
    \begin{tabular}{@{}c|@{}c@{}|@{}c@{}|@{}c@{}|@{}c@{}|@{}c@{}|c@{}c@{}c@{}c|@{}c@{}}
    \hline\hline
        \multirow{3}{*}{Sys.} & \multicolumn{5}{c|}{ Aggregation} & \multicolumn{5}{c}{UASpeech WER\%}  \\ 
    \cline{2-11}
         & \multirow{2}{*}{Method} & \multicolumn{4}{c|}{SI Part} & \multicolumn{4}{c|}{Speech Intelligibility} & \multirow{2}{*}{All} \\
    \cline{3-10}
       & & 1:3 & 1:6 & 1:12 & 1:18 & VL & L & M & H & \\ 
    \hline\hline
        0a & \multicolumn{5}{c|}{centralized SI} & 64.03 & 34.89 & 21.37 & 6.05 & 28.87  \\
    \hline
        0b & \multicolumn{5}{c|}{\tim{centralized SD (adaptor)}} & 59.52 & 33.25 & 22.44 & 6.32 & 27.67  \\ 
    \hline\hline
      \rowcolor{gray!25} 1 & \multicolumn{5}{c|}{-} & 71.50 & 38.22 & 22.50 & 6.19 & 31.45 \\ 

    \hline\hline
        2 & \multirow{4}{*}{\makecell{para.}} & \checkmark & & & & 70.10$^{\dag}$ & 37.54$^{\dag}$ & 21.02$^{\dag}$ & 6.17 & 30.68$^{\dag}$ \\
        3 & & & \checkmark & & & 69.47$^{\dag}$ & 37.64$^{\dag}$ & 21.25$^{\dag}$ & 5.96 & 30.51$^{\dag}$ \\
        4 & & & & \checkmark & & 70.30$^{\dag}$ & 37.71$^{\dag}$ & 21.12$^{\dag}$ & 6.24 & 30.82$^{\dag}$ \\
        5 & & & & & \checkmark & 70.33$^{\dag}$ & 38.20 & 22.17 & 6.25 & 31.12 \\
    \hline\hline
        6 & \multirow{4}{*}{\makecell{embed.}} & \checkmark & & & & 70.55$^{\dag}$ & 37.50$^{\dag}$ & 21.11$^{\dag}$ & 5.91 & 30.71$^{\dag}$ \\
        7 & & & \checkmark & & & 69.10$^{\dag}$ & 37.40$^{\dag}$ & 21.21$^{\dag}$ & 6.18 & 30.46$^{\dag}$ \\
        8 & & & & \checkmark & & 70.12$^{\dag}$ & 37.53$^{\dag}$ & 21.04$^{\dag}$ & 6.19 & 30.70$^{\dag}$ \\
        9 & & & & & \checkmark & 69.95$^{\dag}$ & 38.00 & 22.37 & 6.24 & 31.04 \\ 
    \hline\hline
        \multirow{2}{*}{10} & \multirow{2}{*}{\makecell{para.+\\embed.}} & &\multirow{2}{*}{\checkmark}&&&\multirow{2}{*}{\textbf{69.03}$^{\dag}$} & \multirow{2}{*}{\textbf{37.32}$^{\dag}$} &\multirow{2}{*}{21.22$^{\dag}$} & \multirow{2}{*}{6.22} & \multirow{2}{*}{\textbf{30.43}$^{\dag}$} \\
        &&&&&&&&&\\
    \hline\hline
    \end{tabular}
    \vspace{-0.2cm}
\end{table}

\noindent \textbf{Experiments on the TORGO corpus}: The performance of the two aggregation techniques on TORGO is shown in Table~\ref{tab:torgo_aggregation}. It can be observed that the statistically significant overall WER reductions of up to 0.52\% abs. (4.40\% rel.) and 0.56\% abs. (4.73\% rel.) are obtained by parameter-based \tim{averaging} (Sys.3) and embedding-based \tim{averaging} (Sys.7) over the baseline regularized FedAvg system with quantity-based aggregation strategy(Sys.1), respectively. Combining both leads to
further performance improvement\footnote{trade-off weight is set as 0.3, 0.3,  0.4 for quantity-based, parameter-based and embedding-based \tim{averaging} strategy.}(Sys.10). The improvement brought by the personalization in FL training (0.58\% abs., Sys.10 vs Sys.1) is comparable to that obtained in centralized training settings (0.55\% abs., Sys.0b vs Sys.0a).

\begin{table}[!htbp]
    \centering
    \caption{Performance of parameter (``para.''), embedding (``embed.'')-based \tim{averaging}  for federated learning on \textbf{TORGO}. ``$1$:$3$'' represents that $1^{st}$-$3^{rd}$ transformer layers are chosen as the SI part. ``$1$:$6$'', ``$1$:$12$'', ``$1$:$18$'' have similar meanings. ``Seve.'', ``Mod.'', ``Mild'' are ``severe'', ``moderate'', ``mild'' speech intelligibility. $^{\dag}$ denotes a statistically significant improvement $(\alpha = 0.05)$ obtained over the regularized FedAvg system (Sys.1).}
    \label{tab:torgo_aggregation}
    \vspace{-0.3cm}
    \setlength{\tabcolsep}{1.5pt} 
    \begin{tabular}{@{}c@{}|c|c|c|c|c|ccc|@{}c@{}}
    \hline\hline
        \multirow{3}{*}{Sys.} & \multicolumn{5}{c|}{ Aggregation} & \multicolumn{4}{c}{TORGO WER\%}  \\ 
    \cline{2-10}
         & \multirow{2}{*}{Method} & \multicolumn{4}{c|}{SI Part} & \multicolumn{3}{c|}{Speech Intelligibility} & \multirow{2}{*}{All} \\
    \cline{3-9}
       & & 1:3 & 1:6 & 1:12 & 1:18 & Seve.  & Mod.  & Mild & \\   
    \hline\hline
        0a & \multicolumn{5}{c|}{centralized SI} & 14.43 & 4.90 & 3.10 & 9.36  \\ 
    \hline
        0b & \multicolumn{5}{c|}{\tim{centralized SD (adaptor)}} & 13.38 & 4.80 & 3.17 & 8.81  \\ 
    \hline\hline
      \rowcolor{gray!25} 1 & \multicolumn{5}{c|}{-} & 19.14 & 5.20 & 2.94  & 11.83 \\ 

    \hline\hline
        2 & \multirow{4}{*}{\makecell{para.}} & \checkmark & & & & 18.75 &  4.90& 3.10  & 11.60 \\
        3 & & & \checkmark & & & 18.21$^{\dag}$ & 4.80 & 3.10 & 11.31$^{\dag}$ \\
        4 & & & & \checkmark & & 18.38$^{\dag}$ & 5.31 & 3.02 & 11.48 \\
        5 &  &  &  & & \checkmark & 18.60 & 5.21  & 3.15  & 11.60  \\
        \hline\hline
        6 & \multirow{4}{*}{\makecell{embed.}} & \checkmark & & & & 18.75 & 4.92 & 3.06  & 11.60 \\
        7 & & & \checkmark & & & 18.17$^{\dag}$ & 4.90 & 2.96 & 11.27$^{\dag}$ \\
        8 & & & & \checkmark & & 18.38$^{\dag}$ & 4.92 & 3.17 & 11.45 \\
        9 & & & & & \checkmark & 18.52  & 5.61 & 3.27  &11.69  \\ 
    \hline\hline
        \multirow{2}{*}{10} & \multirow{2}{*}{\makecell{para.+\\embed.}} & &\multirow{2}{*}{\checkmark}&&&\multirow{2}{*}{\textbf{18.02}$^{\dag}$} & \multirow{2}{*}{4.82} &\multirow{2}{*}{3.25} &  \multirow{2}{*}{\textbf{11.25}$^{\dag}$} \\
        &&&&&&&&\\
    \hline\hline
    \end{tabular}
    \vspace{-0.5cm}
\end{table}

\begin{figure}[h!]
    \centering
    \begin{subfigure}[]{0.45\linewidth}
        \centering
        \includegraphics[width=\linewidth]{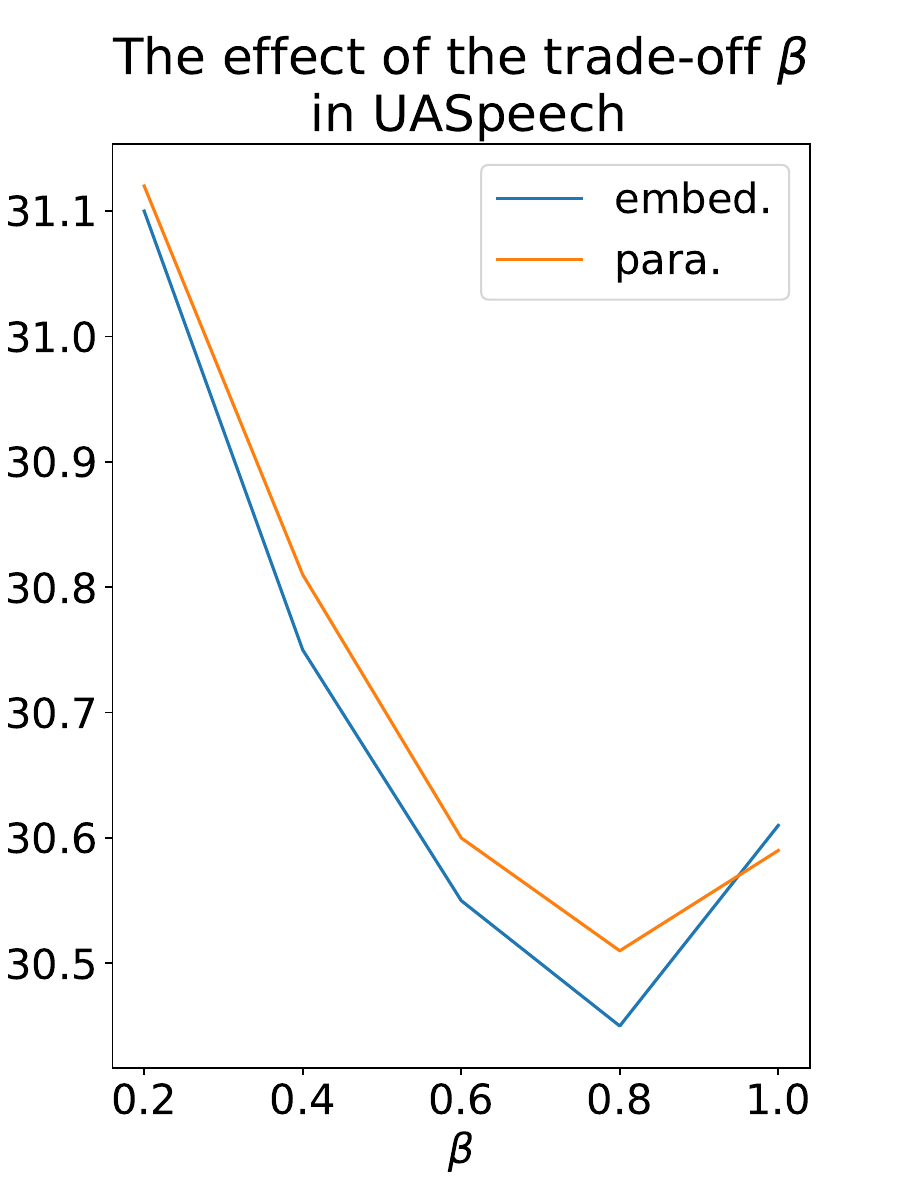}
    \end{subfigure}%
    \hspace{-1em}
    ~ 
    \begin{subfigure}[]{0.45\linewidth}
        \centering
        \includegraphics[width=\linewidth]{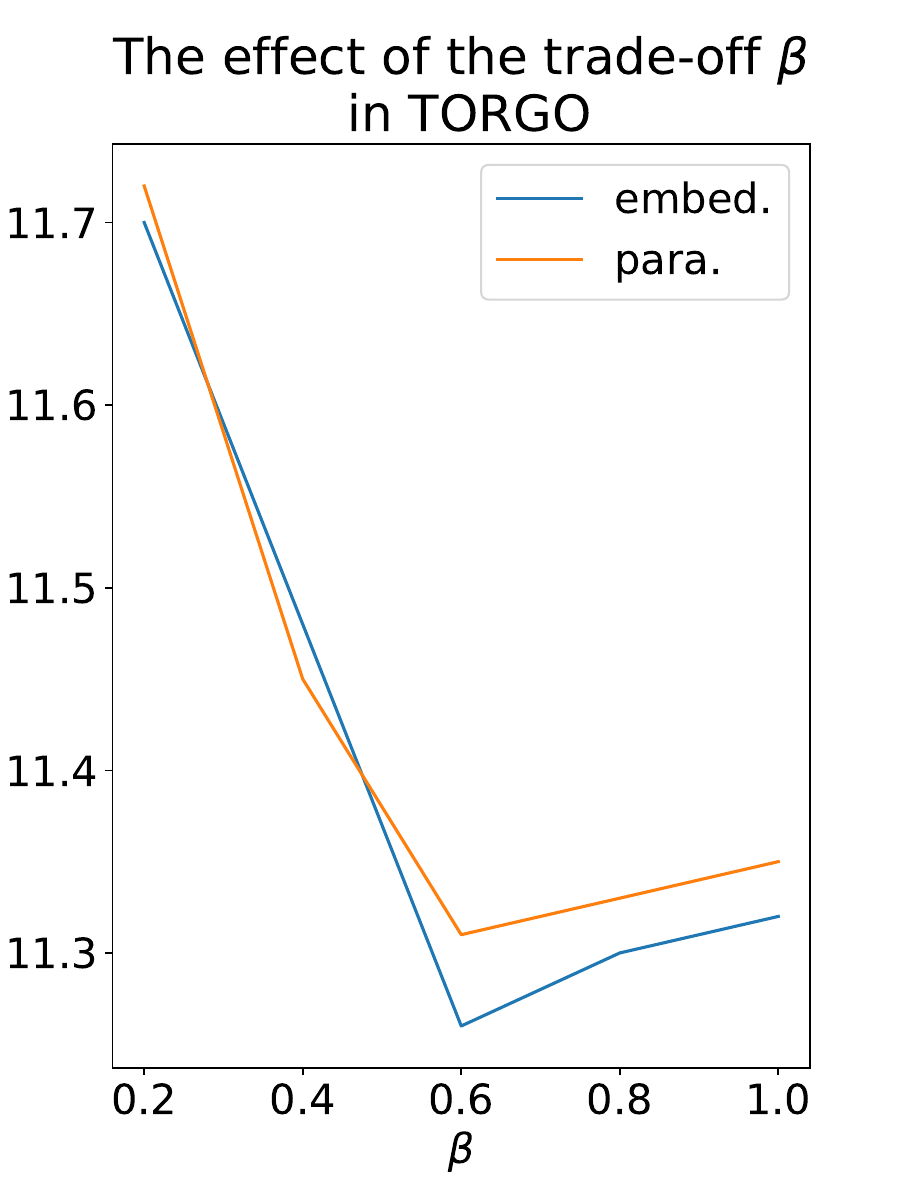}
    \end{subfigure}
    \vspace{-0.5cm}
    \caption{The performance (WER) with different values of the trade-off weight $\beta$ in UASpeech and TORGO.}\label{fig:beta}
    \vspace{-0.3cm}
\end{figure}

\noindent \textbf{Analysis on the effect of the trade-off $\beta$}: Fig.~\ref{fig:beta} shows the impact of the trade-off parameter $\beta$ between the quantity-based and speaker similarity based averaging (Equations~(3) and (5)) when the $1^{st}$-$6^{th}$ transformer layers are chosen as the SI part. For UASpeech, $\beta=0.8$ achieves the smallest WER. For TORGO, WER is smallest when $\beta=0.6$.

\vspace{-0.4cm}

\taozhong{\noindent \textbf{Privacy Analysis}:  Raw audio never leaves the device in our approaches. This is a fundamental privacy advantage over centralized training. Besides, we employ Privacy Amplification~\cite{NEURIPS2018_3b5020bb} via 20\% subsampling and mean-pooling. This temporal collapse ensures the 1024-d embedding is highly resistant to linguistic reconstruction, protecting sensitive user spoken contents while enabling similarity-based personalization.}

\vspace{-0.4cm}
\section{Conclusions}
\vspace{-0.2cm}
This paper explores two aggregation strategies in federated learning to achieve personalization for dysarthric speech recognition. These two strategies include the parameter-based and embedding-based averaging strategies. Experiments on UASpeech and TORGO demonstrate the effectiveness. Future research will focus on aggregation strategies \timtimtim{for} elderly speech.
\label{sec:conclusion}

\taozhong{\section{Acknowledgments}
This research is supported by Hong Kong RGC GRF grant No. 14200021 and 14200324. 
}

\taozhong{
\section{Generative AI Use Disclosure}
Generative AI tools are used for proofreading and grammar correction with minor changes.}

\bibliographystyle{IEEEtran}
\bibliography{new_mybib}

\end{document}